\documentclass[twocolumn,showpacs,preprintnumbers,amsmath,amssymb]{revtex4}

\usepackage{amsmath}
\usepackage{natbib}
\usepackage{graphicx}
\usepackage{epstopdf}
\usepackage{subfigure}
\usepackage{dcolumn}
\usepackage{bm}
\usepackage{color}
\begin{document}

\title{\large Discussion on the equivalence of two relativistic point-particle Lagrangians}

\author{Liubin Wang} 
\author{Xin Wu} 
\email{21200006@sues.edu.cn; wuxin_1134@sina.com}
\affiliation{School of Mathematics, Physics and Statistics,
Shanghai University of Engineering Science, Shanghai 201620,
China}

\begin{abstract}

In 2021, Lei et al. claimed the equivalence between the two
Lagrangians $\mathcal{L}_1
=-mc\sqrt{-g_{\mu\nu}{\dot{x}}^\mu{\dot{x}}^\nu}-V$ and
$\mathcal{L}_2 = \frac{1}{2}mg_{\mu\nu}
{\dot{x}}^\mu{\dot{x}}^\nu-V$ for describing particle dynamics in
combined gravitational and matter fields. In the present work, we
rigorously demonstrate that their equivalence depends critically
on the external potential $V$. Both Lagrangians yield identical
Hamiltonians that strictly satisfy the mass shell constraint, and
are therefore equivalent when $V$ vanishes or corresponds to an
electromagnetic potential. However, they are generally not
equivalent for generic external potentials excluding the
electromagnetic ones. This discrepancy arises because
$\mathcal{L}_1$ and $\mathcal{L}_2$ correspond to different
Hamiltonian formulations. The Hamiltonian derived from
$\mathcal{L}_1$ inherently enforces the mass shell constraint,
whereas the Hamiltonian from $\mathcal{L}_2$ does not. When the
Schwarzschild metric supplemented with an artificial mechanical
potential is taken as a toy model, numerical investigations reveal
that $\mathcal{L}_1$ leads to chaotic behavior, which signifies
non-integrable dynamics. By contrast, $\mathcal{L}_2$ can be shown
analytically to produce integrable dynamics free of chaos. In many
scenarios, $\mathcal{L}_1$ is strongly recommended due to its
theoretical superiority and universality. $\mathcal{L}_2$ is
generally suitable for classical approximate problems involving
low energy and weak gravity. Nevertheless, it is the preferred
choice for strong field problems concerning the dynamics of
charged (or neutral) particles near black holes with (or without)
external electromagnetic fields, owing to its mathematical
simplicity and computational efficiency. Moreover, it can still
satisfy the mass shell constraint when an additional constraint is
imposed on its corresponding Hamiltonian.

\end{abstract}



\maketitle

\section{Introduction}
\label{sec1}

In classical physics [1,2], Lagrangian mechanics provides an
effective approach to unifying the laws of motion for arbitrary
mechanical systems via the Euler-Lagrange equations, which
describe the motion path in configuration space. It avoids the
limitations of Cartesian coordinates and is applicable to a wide
range of complex systems, including rigid bodies, fluids,
gravitational fields, and electromagnetic fields. The Lagrangian
formalism possesses not only covariance but also inherent
connections to symmetry and conservation laws.

Through a Legendre transformation, the Lagrangian formalism can be
converted into the Hamiltonian formulation. Specifically,
Hamiltonian mechanics describes the evolution of a system's point
in phase space by establishing the Hamiltonian canonical
equations. This framework offers a superior perspective for
understanding and analyzing the dynamics of both massive and
massless particles. In particular, it elegantly reveals the
symplectic structure and symplectic symmetry of mechanical systems
within the context of symplectic geometry. This symplectic
structure underpins the construction of symplectic integrators
[3,4]. Symmetry further serves as a cornerstone for identifying
conserved quantities of motion, which determine the integrability
(or non-integrability) of Hamiltonian systems according to
Liouville's theorem [2,5,6]. For non-integrable systems, chaotic
behavior may emerge. While Lagrangian and Hamiltonian mechanics
differ in their mathematical expressions, they are fundamentally
equivalent, as both formalisms represent the same underlying
dynamical systems.

Lagrangian mechanics and Hamiltonian mechanics are related to an
action $S$, which is not an instantaneous state quantity of a
dynamical system but relates to a physical quantity describing the
global properties of this system. Both the Lagrangian and the
Hamiltonian are based on the principle of least action, which is
also known as the variational principle. The variational condition
$\delta S=0$ directly results in the Euler-Lagrange equations and
Hamiltonian canonical equations.

With the aid of the action and variational principle, both
Lagrangian and Hamiltonian mechanics serve as a bridge from
classical physics to modern physics such as quantum physics and
relativistic physics. The Lagrangian in Newtonian mechanics
becomes a Lagrangian density meaning a Lagrangian per spatial unit
volume in relativistic physics. The action representing the time
integral of the Lagrangian in Newtonian mechanics corresponds to
the integral of the Lagrange density over Minkowski spacetime in
relativistic physics. The Einstein field equation can be derived
from the principle of least action with respect to the Lagrangian
density (i.e. Hilbert-Einstein Lagrangian density) of a
gravitational field. It determines the metric tensor $g_{\mu\nu}$
when four coordinate gauge conditions are given.

In addition to the action for the field equation, an action for a
test mass particle freely falling in this gravitational field is
present.  The action is $S_1=-mc^2\int^{2}_{1} ds=\int^{t_2}_{t_1}
\mathcal{L}_1 dt$, where $s$ is the path length of this integral
and the Lagrangian is
\begin{eqnarray}
\mathcal{L}_1=-mc\sqrt{-g_{\mu\nu} \dot{x}^{\mu}\dot{x}^{\nu}}.
\end{eqnarray}
The action is introduced in books on relativity (e.g. [7-10]). The
particle geodesic equation can be derived from the variational
principle of least action $\delta S_1=0$. On the other hand, the
action $S_2=\int^{t_2}_{t_1} \mathcal{L}_2 dt$ with the Lagrangian
\begin{eqnarray}
\mathcal{L}_2=\frac{1}{2}mg_{\mu\nu} \dot{x}^{\mu}\dot{x}^{\nu}
\end{eqnarray}
is sometimes used. When the 4-velocity normalization condition as
an additional constraint is required, the vanishing action
variation $\delta S_2=0$ yields the geodesic equations same as
those derived from the action $S_1$ [10]. In spite of dynamical
equivalence of both Lagrangians, they have their respective
properties including advantages and disadvantages.

Square-root  Lagrangian $\mathcal{L}_1$ has some merits.
\emph{Application scenarios}: $\mathcal{L}_1$ is well applicable
to the motion of massive particles in all regimes from
non-relativistic case to ultra-relativistic case.
\emph{Relativistic covariance}: The Lagrangian satisfies strict
relativistic covariance under coordinate transformations.
\emph{Mass shell constraint}: $\mathcal{L}_1$ inherently endows
the mass shell constraint condition. In this case, the Lagrangian
describes a constrained physical system [55].

Non-square-root Lagrangian $\mathcal{L}_2$ has its advantages as
follows. \emph{Mathematical simplicity}: For timelike geodesics,
the quadratic terms of the 4-velocity in $\mathcal{L}_2$ are
simpler in numerical simulations and analytical perturbative
studies than the square-root term of the 4-velocity in
$\mathcal{L}_1$. \emph{Classical energy concept}: Its Hamiltonian
usually represents total energy conservation in static spacetimes,
and is well correlated with classical mechanics and quantum
mechanics. However, it has theoretical shortcomings. \emph{No
constraints}: It has no mass shell and Hamiltonian constraint
conditions in general unless an additional constraint is given to
its corresponding Hamiltonian. \emph{Lack of relativistic
covariance}: $\mathcal{L}_2$ violates the strict covariance
principle of general relativity in the description of geodesic
equations  and spacetime geometry. \emph{Application limitations}:
It is suitable for the weak-field and low-speed limits, but
becomes generally useless for strong gravitational fields. See
several books [7-10] on general relativity for more details
regarding the properties of both Lagrangians.

If an extra matter source is included, the total action for the
gravitational and mater fields is the sum of both the
gravitational action and the matter action. The field equations
are obtained via the total action. On the other hand, the action
for a test mass particle interacting with the gravitational field
given by the metric $g_{\mu\nu}$ and the matter field described by
an external potential $V$ (such as an electromagnetic potential or
a gravitational potential) corresponds to the Lagrangian
\begin{eqnarray}
\mathcal{L}_1=-mc\sqrt{-g_{\mu\nu} \dot{x}^{\mu}\dot{x}^{\nu}}-V.
\end{eqnarray}
This Lagrangian can provide the particle motion equations. It has
been used to study the motion of massive or massless test
particles in the literature [11-17]. There is a non-square-root
Lagrangian
\begin{eqnarray}
\mathcal{L}_2=\frac{1}{2}mg_{\mu\nu} \dot{x}^{\mu}\dot{x}^{\nu}-V.
\end{eqnarray}
Using the non-square-root Lagrangian, many authors have focused on
the regular and chaotic dynamics of charged particles near black
holes immersed in external magnetic fields [18-39].

Letting the external potential be the $t$ component of an
electromagnetic potential
\begin{eqnarray}
V=qA_t(r)\dot{t},
\end{eqnarray}
where $q$ denotes the particle charge and $A_t(r)$ is a function
of $r$, Lei et al. [13] explored particle motion near a black hole
with the external  electromagnetic field. In their paper, Eqs.
(C3) and (C14) with respect to the two Lagrangian formalisms are
\begin{eqnarray}
\mathcal{L}_1 &=&-mc\sqrt{-g_{\mu\nu}{\dot{x}}^\mu{\dot{x}}^\nu}-qA_t(r)\dot{t}, \\
\mathcal{L}_2 &=&
\frac{1}{2}mg_{\mu\nu}{\dot{x}}^\mu{\dot{x}}^\nu-qA_t(r)\dot{t}.
\end{eqnarray}
They found that both  Lagrangian formalisms produce the same value
of the maximal Lyapunov exponent for the particle radial motions
at the equilibrium position. Considering this fact and the result
on the equivalence of both  Lagrangians in Eqs. (1) and (2), they
claimed in terms of their Eq. (18) and Footnote 1 that the two
Lagrangian formalisms $\mathcal{L}_1$ and $\mathcal{L}_2$ of Eqs.
(3) and (4) are dynamically equivalent for the general external
potential
\begin{eqnarray}
V=\psi(x^i),
\end{eqnarray}
where $\psi$ is a function of coordinates $x^i$.

Is the equivalence of both Lagrangians $\mathcal{L}_1$ and
$\mathcal{L}_2$ in Eqs. (3) and (4) always true for any external
potential $V(x^i,\dot{x}^\mu)$? In the present work, no
Euler-Lagrange equations but Hamiltonian formalisms will be
considered to answer this question. This consideration is based on
the above-mentioned advantages of Hamiltonian formalisms.
Hamiltonian dynamics are independent of the choice of spacetime
coordinates including time parameter. The Hamiltonian systems
intuitively show integrals of motion involving the  mass shell
constraint. The integrals of motion are useful to determine
integrability or non-integrability of Hamiltonian systems through
the integrability criterion of Liouville's theorem [2,5,6]. In
addition to the main question, other questions are present. If
both Lagrangians are not always equivalent, which of them is more
physically preferable for the description of particle motion? When
a massless particle (such as a photon) interacting with the
external potential $V$, which of them is suitably applied?
$\mathcal{L}_2$ has its limitations that it generally loses the
mass shell constraint and is generally inappropriate for strong
fields. However, a large body of studies [18-39] have focused on
the application of $\mathcal{L}_2$ to the dynamics of charged
particles near black holes with external magnetic fields. Are
these studies physically tenable from the relativistic
perspective? These questions will be addressed in this paper.

The rest of this paper is organized as follows. In Sect. 2, we
theoretically analyze the dynamical equivalence or nonequivalence
of both Lagrangians on the basis of three cases of the external
potentials. A toy model is used to check our analytical results in
Sect. 3. Several remarks are given to the two Lagrangians in Sect.
4. Finally, the main conclusions are drawn in Sect. 5.

\section{Theoretical analysis of both Lagrangian formulations} \label{sec2}

Suppose that the metric $g_{\mu\nu}$ of Eqs. (3) and (4) is
static, axisymmetric in coordinates $x^{\mu}=(t,r,\theta,\phi)$.
In addition, the external potential $V(x^i,\dot{x}^\mu)$
($i=1,2,3$ and $\mu=0,1,2,3$) is axisymmetric. $m$ is the particle
mass. For simplicity, we set $g_{0i}=0$. The speed of light $c$
and the gravity constant $G$ are taken in geometric units as
$c=G=1$. The dot in the above equations denotes the derivative for
an arbitrary parameter $\lambda$.

Under the above assumptions,  both Lagrangians $\mathcal{L}_1$ and
$\mathcal{L}_2$ of Eqs. (3) and (4) have the conserved particle
energies $E$ and $\mathcal{E}$, which correspond to generalized
momenta along the time $t$ axis:
\begin{eqnarray}
-E&=&p_t=\frac{\partial\mathcal{L}_1}{\partial
\dot{t}}=\frac{mg_{00} \dot{t}}{\sqrt{-g_{\mu\nu}
{\dot{x}}^\mu{\dot{x}}^\nu}}-\frac{\partial V}{\partial \dot{t}}, \\
-\mathcal{E}&=&P_t=\frac{\partial\mathcal{L}_2}{\partial
\dot{t}}=mg_{00}\dot{t}-\frac{\partial V}{\partial \dot{t}}.
\end{eqnarray}
There are also the conserved particle angular momenta $L$ and
$\Phi$, which correspond to generalized momenta along the $\phi$
direction:
\begin{eqnarray}
L &=&p_\phi=\frac{\partial\mathcal{L}_1}{\partial
\dot{\phi}}=\frac{mg_{33}\dot{\phi}}{\sqrt{-g_{\mu\nu}
{\dot{x}}^\mu{\dot{x}}^\nu}}-\frac{\partial V}{\partial \dot{\phi}}, \\
\Phi &=& P_\phi=\frac{\partial\mathcal{L}_2}{\partial
\dot{\phi}}=mg_{33}\dot{\phi}-\frac{\partial V}{\partial
\dot{\phi}}.
\end{eqnarray}

In what follows, we theoretically explore whether the two
Lagrangian formalisms in Eqs. (3) and (4) are equivalent according
to three cases of the external potential $V$.

\subsection{Electromagnetic potential}

Assume that the external potential is an electromagnetic vector
potential $A_\mu(r,\theta)=(A_t, A_i)$ for the description of
Maxwell field
\begin{equation}
V_I(x^i,\dot{x}^\mu)=qA_\mu(r,\theta)\dot{x}^\mu.
\end{equation}
Some examples of electromagnetic potentials are self-consistent
electromagnetic potentials [18,31], Wald electromagnetic
potentials [40,41] and generalized electromagnetic potentials
[30,31,42-44]. Eq. (5) is only a component of the electromagnetic
vector potential $A$. In this case, the Lagrangian forms are
written as
\begin{eqnarray}
\mathcal{L}_1 &=&-m\sqrt{-g_{00}\dot{t}^2-g_{ij}
{\dot{x}}^i{\dot{x}}^j}-qA_t\dot{t}-qA_i\dot{x}^i, \\
\mathcal{L}_2 &=&
\frac{1}{2}mg_{\mu\nu}{\dot{x}}^\mu{\dot{x}}^\nu-qA_\mu\dot{x}^\mu.
\end{eqnarray}
The expression for $\mathcal{L}_1$ appears in several works (e.g.,
Refs. [12,14]). $\mathcal{L}_2$ has been widely adopted in the
literature (e.g., Refs. [18-39]).

For the Lagrangian $\mathcal{L}_1$ of Eq. (14), the coordinates
$x^\mu$ correspond to their canonical generalized momenta
\begin{eqnarray}
p_t &=& \frac{\partial \mathcal{L}_1}{\partial
\dot{t}}=\frac{mg_{00}}{\sqrt{-g_{00}\dot{t}^2-g_{ij}{\dot{x}}^i{\dot{x}}^j}}-qA_t=-E,
\\
p_i &=& \frac{\partial \mathcal{L}_1}{\partial \dot{x}^i}
=\frac{mg_{ij}\dot{x}^j}{\sqrt{-g_{00}\dot{t}^2-g_{ij}{\dot{x}}^i{\dot{x}}^j}}-qA_i.
\end{eqnarray}
It is easy to check that the generalized momentum $p_{\mu}$ has
the relation $p_{\mu}p^{\mu}\neq-m^2$, but for timelike motion the
readjusted (or redefined) momentum
$\tilde{p}_{\mu}=p_{\mu}+qA_\mu$ naturally satisfies the
constraint condition
\begin{equation}
\tilde{p}_{\mu}\tilde{p}^{\mu}=g^{00}\tilde{p}^2_{t}+g^{ij}\tilde{p}_{i}\tilde{p}_{j}
=-m^2.
\end{equation}
This condition is called as the  mass shell constraint belonging
to a first-class constraint [55], which arises due to
reparametrization invariance (gauge invariance) of the action.

In the generic parametrization with parameter $\lambda$, the
canonical Hamiltonian $H_c = p_\mu\dot{x}^\mu-\mathcal{L}_1$ from
the Legendre transformation of the Lagrangian $\mathcal{L}_1$ in
Eq. (14) is identical to zero, i.e., $H_c =0$. The result is just
that of the textbook [55]. As shown in [55], the vanishing
Hamiltonian is obtained  because the Lagrangian is singular and
homogeneous of degree one in the velocities. Consequently, the
momentum $p_\mu$ is homogeneous of degree zero in the velocities
and there is no unique solution with respect to velocity
$\dot{x}^\mu$. Dirac's methods can be used to define the
Hamiltonian system satisfying the constraint (18). This is
\emph{the first path} for obtaining the constrained Hamiltonian
dynamics of $\mathcal{L}_1$. \emph{A second path} with the use of
the static gauge $\lambda=t$ relies on the Legendre transformation
and arrives at the Hamiltonian
\begin{eqnarray}
H_1 &=& p_i\dot{x}^i-\mathcal{L}_1 \nonumber
\\
&=&\sqrt{-g_{00}}\sqrt{m^2+(p_i+qA_i)(p^i+qA^i)} +qA_t.
\end{eqnarray}
Note that $p_t$ is undefined in the static gauge with $\dot{t}=1$
since the derivative for a constant cannot be computed. In spite
of this, no $p_t$ in Eq. (16) but $p_i$ in Eq. (17) is used in the
derivation of Eq. (19). If the constraint of Eq. (16) is adjoined
to $p_t$, that is, the first term of Eq. (18) in the static gauge
is defined as $g^{00}\tilde{p}^2_{t}=m^2g_{00}
/(-g_{00}-g_{ij}\dot{x}^i\dot{x}^j)$, then the constraint (18) is
still satisfied. In addition, $H_1=E$ is easily derived. This
derivation is given here. Considering that $H_c=0$, $\dot{t}=1$
and $p_t=-E$, we have $H_1=H_c-\dot{t}p_t=0-1*(-E)=E$. When $q=0$
is given and $\sqrt{-g_{00}}$ is eliminated, the Hamiltonian of
Eq. (19) is consistent with Eq. (2.18) of [55]. In fact, the
Hamiltonian (19) can completely become Eq. (2.18) of [55] if our
parameter $\lambda$ is readjusted as
$d\lambda=dt_c/\sqrt{-g_{00}}$, where $t_c$ is the parameter of
Eq. (2.18) in [55]. \emph{A third path} that uses $\lambda$ as a
parameter but does not adopt the static gauge $\lambda=t$ is the
expression of energy $E$ taken as the Hamiltonian $H_1$, where $E$
is exactly solved from the readjusted constraint (18). It is clear
that  the Hamiltonian is only a variant of the constraint (18) and
naturally contains the consistency of the constraint. In other
words, the Hamiltonian obtained via the third path is identical to
that derived from the second path. This fully demonstrates that
the Hamiltonian from the second path with the use of the static
gauge accurately satisfies the mass shell constraint (18). The
three paths are independent but have close relations in the
derivation of $H_1$. Paths 1 and 2 lead to $H_1=E$, and Path 3
obtains the expression of $E$ from the constraint (18).

For the Lagrangian $\mathcal{L}_2$ of Eq. (15), the generic
parameter $\lambda$ is still adopted and the canonical generalized
momenta are defined by
\begin{equation}
P_\mu=\frac{\partial \mathcal{L}_2}{\partial \dot{x}^\mu}
=mg_{\mu\nu}\dot{x}^\nu-qA_\mu.
\end{equation}
The particle has its conserved energy
\begin{equation}
\mathcal{E}=-P_t=-mg_{t\nu}\dot{x}^\nu+qA_t.
\end{equation}
A Hamiltonian corresponding to the Lagrangian $\mathcal{L}_2$ is
expressed as
\begin{eqnarray}
K &=& P_\mu\dot{x}^\mu-\mathcal{L}_2 \nonumber
\\
&=& \frac{1}{2m}g^{\mu\nu}(P_\mu+qA_\mu)(P_\nu+qA_\nu)\nonumber
\\
&=& \frac{1}{2m}g^{00}(\mathcal{E}
-qA_t)^2+\frac{1}{2m}g^{ij}(P_i+qA_i)
\nonumber \\
&& \cdot(P_j+qA_j).
\end{eqnarray}
The Hamiltonian $K$ does not explicitly depend on the parameter
$\lambda$  and therefore is always a conserved quantity. For the
timelike motion, this conserved quantity is artificially chosen as
\begin{equation}
K=-\frac{1}{2}m.
\end{equation}
In fact, this artificial choice is so good that the Hamiltonian of
Eq. (22) consistently possesses the mass shell constraint
\begin{equation}
\tilde{P}_{\mu}\tilde{P}^{\mu}=2mK=-m^2,
\end{equation}
where $\tilde{P}_{\mu}=P_{\mu}+qA_\mu$ is a readjusted momentum
with respect to the momentum (20). Although the momentum
$\tilde{p}_{\mu}$ for $\mathcal{L}_1$ and the momentum
$\tilde{P}_{\mu}$ for $\mathcal{L}_2$ satisfy the same constraint
condition (18) or (24), they have different mechanisms in the
momentum conservations. The momentum $\tilde{p}_{\mu}$ inherently,
accurately reflects the relation (18) in the Hamiltonian $H_1$,
while the momentum $\tilde{P}_{\mu}$ with the additional
constraint condition (24) is subject to the attached condition
(23). If $K$ is not constrained by Eq. (23), $\tilde{P}_{\mu}$
must fail to have the relation (24). In other words,
$\tilde{p}_{\mu}$ self-consistently contains the mass shell
constraint (18), whereas $\tilde{P}_{\mu}$ is fit for the mass
shell constraint (24) only when the additional condition (23) is
given to the Hamiltonian $K$. Noting that Eqs. (22) and (23) are
the readjusted constraint (24), we solve the energy from  the
constraint:
\begin{equation}
\mathcal{E} =qA_t+\sqrt{-g_{00}}\sqrt{m^2+g^{ij}
(P_i+qA_i)(P_j+qA_j)}.
\end{equation}
This energy is taken as a new Hamiltonian system
\begin{eqnarray}
H_2 =\mathcal{E}.
\end{eqnarray}
Clearly, the Hamiltonians $H_2$ and $H_1$ have the relation
$H_2=H_1$ if $p_i$ and $P_i$ are regarded as the same notations.
Therefore, the Hamiltonian $H_2$ like $H_1$ consistently possesses
the mass shell constraint (24) when the additional constraint (23)
is given to the Hamiltonian $K$ from the Legendre transformation
of $\mathcal{L}_2$.

In fact, the new Hamiltonian $H_2$ is equivalent to the
Hamiltonian $K$. The truth of this proposition is proved as
follows. Taking the time $\lambda$ as a coordinate and its
canonical momentum as $p_0=-H_2=-\mathcal{E}$, we have an extended
phase space Hamiltonian
\begin{eqnarray}
H^*_2 &=& H_2+p_0 \nonumber \\
&=& \sqrt{-g_{00}}\sqrt{m^2+g^{ij}(p_i+qA_i)(p_j+qA_j)} \nonumber
\\ && +p_0+qA_t.
\end{eqnarray}
This Hamiltonian is always identical to zero for any time
$\lambda$, $H^*_2=0$. Consider that a time transformation is
\begin{eqnarray}
d\lambda = Fd\delta,
\end{eqnarray}
where $F$ is a time transformation function
\begin{eqnarray}
F &=& [\sqrt{-g_{00}}\sqrt{m^2+g^{ij} (p_i+qA_i)(p_j+qA_j)}
\nonumber \\ && -(p_0+qA_t)]/(-2g_{00}).
\end{eqnarray}
In the new time $\delta$, a time-transformed Hamiltonian with
respect to the Hamiltonian $H^*_2$ is
\begin{eqnarray}
\Gamma= FH^*_2=K+\frac{m}{2}.
\end{eqnarray}
This shows that only one difference between the Hamiltonians $K$
of Eq. (22) and $H_2$ of Eq. (26) is time scales (or time
parameters). When the time parameter is re-scaled or readjusted
according to Eq. (28) with Eq. (29), the Hamiltonian $H_2$ can be
converted back to the Hamiltonian $K$. In essence, both $H_2$ and
$K$ are equivalent. The Hamiltonian remains invariant under time
parameter transformations. The equivalence provides a theoretical
basis for the application of time transformation methods to
designing explicit symplectic integrators with adaptive time-step
controls for the Hamiltonian $K$ of Eq. (22) with the additional
constraint of Eq. (23) [45-47]. Such adaptive time-step algorithms
are relatively suitable for treating highly eccentric orbits and
close encounters between objects in the solar system. The time
transformation methods  also play an important role in
successfully designing adaptive time-step explicit symplectic
integrators for many black hole spacetimes [48-50].

As analytically demonstrated above,  both Hamiltonians $H_1$ and
$H_2$ are the particle energies $E$ and $\mathcal{E}$, which are
derived from the mass shell constraints. In fact, the Hamiltonian
$H_2$ is the same as the Hamiltonian $H_1$ through the time
transformation. This sufficiently shows the equivalence of the two
Lagrangian formalisms of Eqs. (14) and (15) when the external
potential $V$ is the electromagnetic potential. The fundamental
reason for the equivalence of both Lagrangians is that the
Hamiltonian systems of both Lagrangians self-consistently possess
the mass shell constraints. Although the external electromagnetic
field in Eq. (13) is too weak to change the spacetime geometry, it
can exert an important influence on charged particle motion. Even
chaotic dynamics of charged particles are allowed in either the
Lagrangian formalism $\mathcal{L}_1$ (corresponding to the
Hamiltonian formalism $H_1$ of Eq. (19)) or the Lagrangian
formalism $\mathcal{L}_2$ (corresponding to the Hamiltonian
formalism $K$ of Eq. (22) with the additional constraint of Eq.
(23)) (see e.g. [19-24,29-34]).

For the potential of Eq. (5) as a special case of the
electromagnetic potential in Eq. (13), the two Lagrangian
formalisms of Eqs. (6) and (7) are naturally equivalent. This is
why the authors of [13] obtained the same expression of the
maximal Lyapunov exponent for the only radial motion of charged
particles at the equilibrium position in these two Lagrangian
forms. When $q=0$ (corresponding to the vanishing external
potential), the two Lagrangian forms are those of Eqs. (1) and
(2), which must be equivalent due to the conservation of the mass
shell constraints.

\subsection{Non-electromagnetic potential}

Now, the external potential is the non-electromagnetic potential
of Eq. (8),  which is rewritten as
\begin{equation}
V_{II}=\psi(r,\theta).
\end{equation}
Unlike the electromagnetic potential, $\psi$  is a generic
potential such as a mechanical potential or an effective
potential. Eqs. (3) and (4) are of the forms
\begin{eqnarray}
\mathcal{L}_1 &=&-m\sqrt{-g_{00}\dot{t}^2-g_{ij}
{\dot{x}}^i{\dot{x}}^j}-\psi(r,\theta), \\
\mathcal{L}_2 &=& \frac{1}{2}mg_{\mu\nu}\dot{x}^\mu
\dot{x}^\nu-\psi(r,\theta).
\end{eqnarray}
Here, $\dot{x}^\mu$ is still the derivative of coordinate $x^\mu$
with respect to an arbitrary parameter $\lambda$.

The Lagrangian $\mathcal{L}_1$ of Eq. (32) corresponds to its
canonical generalized momenta
\begin{eqnarray}
p_t&=& \frac{\partial \mathcal{L}_1}{\partial
\dot{t}}=\frac{mg_{00}}{\sqrt{-g_{00}\dot{t}^2-g_{ij}{\dot{x}}^i{\dot{x}}^j}}=-E,
\\
p_i &=& \frac{\partial \mathcal{L}_1}{\partial \dot{x}^i}
=\frac{mg_{ij}\dot{x}^j}{\sqrt{-g_{00}\dot{t}^2-g_{ij}{\dot{x}}^i{\dot{x}}^j}}.
\end{eqnarray}
Obviously, the generalized momentum $p_{\mu}$ for timelike motion
coherently entails the dispersion relation [15] or the mass shell
constraint
\begin{equation}
p_{\mu}p^{\mu}=g^{00}E^2+p_ip^i=-m^2.
\end{equation}
After the energy $E$ is solved from the above equation, the
Lagrangian $\mathcal{L}_1$ corresponds to the Hamiltonian
\begin{eqnarray}
H_1 =E+\psi=\sqrt{-g_{00}}\sqrt{m^2+p_ip^i}+\psi.
\end{eqnarray}
The Hamiltonian from the third path is consistent with that
obtained from the second path (corresponding to Legendre
transformation of $\mathcal{L}_1$). The first term
$E=\sqrt{-g_{00}}\sqrt{m^2+p_ip^i}$ corresponds to the particle
energy from the gravitational field, and the second term $\psi$ is
the particle energy  from the matter field. Due to the energy
exchange, the first term does not remain constant although the
Hamiltonian quantity is conserved. Clearly, the mass shell
constraint is always satisfied in the Hamiltonian $H_1$,
regardless of how the first term changes.

For the Lagrangian $\mathcal{L}_2$ of Eq. (33), the canonical
generalized momentum is
\begin{equation}
P_\mu=\frac{\partial \mathcal{L}_2}{\partial \dot{x}^\mu}
=mg_{\mu\nu}\dot{x}^\nu.
\end{equation}
Eq. (10) for the conserved particle energy is
\begin{equation}
\mathcal{E}_g=-P_t=-mg_{t\nu}\dot{x}^\nu.
\end{equation}
The Hamiltonian derived from the Legendre transformation of
$\mathcal{L}_2$ is written as
\begin{eqnarray}
K=\frac{1}{2m}g^{\mu\nu}P_\mu P_\nu+\psi.
\end{eqnarray}
If $K$ is still given by the additional constraint condition (23),
we have
\begin{eqnarray}
P_\mu P^\mu=g^{\mu\nu}P_\mu P_\nu=2m(K-\psi)=-m^2-2m\psi.
\end{eqnarray}
This leads to the particle energy
\begin{eqnarray}
\mathcal{E}=\sqrt{-g_{00}}\sqrt{-2mK+2m\psi+P_iP^i}.
\end{eqnarray}
The two energies $\mathcal{E}$ and  $\mathcal{E}_g$ are different.
$\mathcal{E}$ is the energy for the particle interacting with the
gravitational and matter fields, and $\mathcal{E}_g$ is only the
particle energy from the gravitational field. The energy
$\mathcal{E}$ is regarded as a Hamiltonian
\begin{eqnarray}
H_2= \mathcal{E}=\sqrt{-g_{00}}\sqrt{m^2+2m\psi+P_iP^i}.
\end{eqnarray}
Note that $\mathcal{E}$ is derived from Eq. (41) but is not
derived from the following mass shell constraint condition
\begin{equation}
P_\mu P^\mu= -m^2.
\end{equation}
Therefore, the mass shell constraint is not kept in the
Hamiltonian $H_2$ corresponding to the Lagrangian $\mathcal{L}_2$
for the case of $\psi\neq 0$.

If the momentum $P_\mu$ rather than the Hamiltonian $K$ is given
the additional constraint condition (44), then Eq. (40) becomes
$K=\frac{1}{2m}g^{\mu\nu}P_\mu P_\nu+\psi=-\frac{m}{2}+\psi$. This
shows no exchange between the particle's gravity energy and the
particle's matter energy. The absence of energy exchange is
inconsistent with physical reality. On the other hand, if
$\mathcal{E}_g$ solved from the mass shell constraint (44) is used
to obtain the Hamiltonian $\bar{H}_2=\psi+\mathcal{E}_g$, then
$\bar{H}_2$ and $H_1$ are the same. However, there is a fatal
problem that $\bar{H}_2$ is incompatible with the Hamiltonian
derived from the Legendre transformation of $\mathcal{L}_2$.

It is clearly shown via the above demonstrations that the two
Hamiltonians in Eqs. (37) and (43) have different expressions for
the external potential $\psi\neq0$. They are based on the particle
energies corresponding to the generalized momenta along the time
axis: $H_1=-p_t+\psi=E+\psi$ and $H_2=-P_t=\mathcal{E}$, where $E$
is exactly solved from the mass shell constraint (36) but
$\mathcal{E}$ satisfies a modified form (41) of the Hamiltonian
$K$ rather than the mass shell constraint (44). Thus, the two
Lagrangians of Eqs. (32) and (33) are not dynamically equivalent.

\subsection{Combination of two potentials}

Let the external potential be the superposition of the
electromagnetic potential and the other generic external
potential:
\begin{equation}
V_{III}=\varphi+\psi=qA_\mu(r,\theta)\dot{x}^\mu+\psi(r,\theta).
\end{equation}
Eqs. (3) and (4) are reexpressed as
\begin{eqnarray}
\mathcal{L}_1 &=&-m\sqrt{-g_{00}\dot{t}^2-g_{ij}
{\dot{x}}^i{\dot{x}}^j}-qA_t\dot{t} \nonumber \\
&& -qA_i\dot{x}^i-\psi(r,\theta), \\
\mathcal{L}_2 &=&
\frac{1}{2}mg_{\mu\nu}{\dot{x}}^\mu{\dot{x}}^\nu-qA_\mu\dot{x}^\mu-\psi(r,\theta).
\end{eqnarray}
Here, $\lambda$ is still an arbitrary parameter and $\dot{x}^\mu$
is defined as $\dot{x}^\mu=\frac{dx^\mu}{d\lambda}$.

The generalized momenta $p_\mu$ of $\mathcal{L}_1$ are still those
of Eqs. (16) and (17). The readjusted momenta $\tilde{p}_\mu$ also
satisfy the mass shell constraint condition (18). The Hamiltonian
obtained through the Legendre transformation to $\mathcal{L}_1$ is
of the form
\begin{eqnarray}
H_1 &=& E+\psi \nonumber
\\ &=& \sqrt{-g_{00}}\sqrt{m^2+(p_i+qA_i)(p^j+qA^j)} \nonumber
\\ && +qA_t+\psi,
\end{eqnarray}
where $E$ as the energy solution of the mass shell constraint (18)
is still obtained from the third path. $\mathcal{L}_2$ has its
generalized momenta $P_\mu$, which are of the form (20). The
Legendre transformation to $\mathcal{L}_2$ has the Hamiltonian
\begin{eqnarray}
K &=& \frac{1}{2m}g^{\mu\nu}(P_\mu+qA_\mu)(P_\nu+qA_\nu)+\psi
\nonumber
\\
&=&
\frac{1}{2m}g^{00}(\mathcal{E}-qA_t)^2+\frac{1}{2m}g^{ij}(P_i+qA_i)
\nonumber \\
&& \cdot(P_j+qA_j)+\psi.
\end{eqnarray}
When $K$ is given by the additional constraint (23), we obtain the
particle energy $\mathcal{E}$ as a Hamiltonian
\begin{eqnarray}
H_2 &=&
\mathcal{E}=\sqrt{-g_{00}}\sqrt{m^2+2m\psi+(P_i+qA_i)(P^j+qA^j)}
\nonumber
\\ && +qA_t.
\end{eqnarray}
Although the additional constraint (23) to $K$ is required, the
readjusted momenta $\tilde{P}_\mu$ do not satisfy the mass shell
constraint condition (24). That is, the readjusted momenta have
the relation
\begin{eqnarray}
\tilde{P}_\mu \tilde{P}^\mu\neq-m^2.
\end{eqnarray}
Because $H_1$ and $H_2$ are two distinct Hamiltonians for
$\psi\neq0$, $\mathcal{L}_1 $ and $\mathcal{L}_2$ represent two
different dynamical systems.

In short, the above analytical methods clearly demonstrate that
the two Lagrangian forms of Eqs. (3) and (4) are equivalent in the
dynamics when the external potential is zero or the
electromagnetic potential. The fundamental reason is that they can
keep the mass shell constraints and their corresponding
Hamiltonians have the same expressions. However, both Lagrangians
are generally nonequivalent for the other external potentials.
This is because they have different Hamiltonian expressions. The
mass shell constraint can be satisfied in the Hamiltonian $H_1$ of
$\mathcal{L}_1 $, while is difficultly inserted into the
Hamiltonian $K$ or $H_2$ of $\mathcal{L}_2$. These analytical
results are obtained on the basis of the assumption $g_{0i}=0$.
They should also be correct if $g_{0i}\neq0$. In this case, the
related derivations become more complicated.

\section{A toy model to examine dynamical
differences between $\mathcal{L}_1$ and $\mathcal{L}_2$}
\label{sec3}

We take $g_{\mu\nu}$ as the metric tensor of Schwarzschild black
hole:
\begin{eqnarray}
g_{tt} &=& -(1-\frac{2}{r}), ~~~~~ g_{rr}=\frac{1}{1-\frac{2}{r}},
\nonumber
\\ g_{\theta\theta} &=& r^2, ~~~~~~~~~~~~~~ g_{\phi\phi} =
r^2\sin^2\theta.
\end{eqnarray}
This metric has its contravariant nonzero components
\begin{eqnarray}
g^{tt} &=& -\frac{1}{1-\frac{2}{r}}, ~~~~~~~
g^{rr}=(1-\frac{2}{r}), \nonumber
\\ g^{\theta\theta} &=& \frac{1}{r^2}, ~~~~~~~~~~~~~ g^{\phi\phi} =\frac{1}{
r^2\sin^2\theta}.
\end{eqnarray}
The black hole mass is taken as one geometric unit, $M=1$. The
event horizon is $r_g=2$. Set the external potential as an
artificially constructed mechanical potential unlike the
electromagnetic potential:
\begin{eqnarray}
V(r,\theta)=\psi_1=\frac{\omega}{r^2}[(r-r_c)^2+r^2_g\theta^2],
\end{eqnarray}
where $r_c$ is a parameter representing the center of the harmonic
potential $\tilde{V}=(r-r_c)^2+4\theta^2$ and $\omega$ is another
parameter. For the dynamics of massive particles, $m=1$ is taken.

\subsection{Integrable dynamics of the Lagrangian $\mathcal{L}_2$ }

In terms of Eq. (33) with Eqs. (52)-(54), the Lagrangian
$\mathcal{L}_2$ is of the form
\begin{eqnarray}
\mathcal{L}_2 &=&
\frac{1}{2}[-(1-\frac{2}{r})\dot{t}^2+\frac{\dot{r}^2}{1-\frac{2}{r}}
+r^2\dot{\theta}^2  \nonumber \\ && +r^2\sin^2\theta\dot{\phi}^2]
-\frac{\omega}{r^2}[(r-r_c)^2+4\theta^2].
\end{eqnarray}
Its  corresponding Hamiltonian is Eq. (40) with $P\rightarrow p$:
\begin{eqnarray}
K &=&
-\frac{\mathcal{E}^2}{2(1-\frac{2}{r})}+\frac{p^2_r}{2}(1-\frac{2}{r})
+ \frac{p^2_\theta}{2r^2} + \frac{\Phi^2}{2r^2\sin^2\theta}
\nonumber
\\ && + \frac{\omega}{r^2} [(r -r_c)^2 + 4\theta^2].
\end{eqnarray}
Note that this Hamiltonian is not the Hamiltonian of Eq. (43).
Even if the Hamiltonian $K$ is still given by Eq. (23), it loses
the mass shell constraint (44).

It is easy to find that the variables of Eq. (56) can be separated
into two terms:
\begin{eqnarray}
&& Kr^2 +\frac{r^3\mathcal{E}^2}{2(r-2)}- \frac{r}{2}
(r-2) p_r^2-\omega(r -r_c)^2 \nonumber \\
&&=\frac{p_\theta^2}{2}+ \frac{\Phi^2}{2\sin^2 \theta} + 4\omega
\theta^2=Q.
\end{eqnarray}
In fact, this equation arises from the Hamilton-Jacobi equation.
This shows the existence of a fourth conserved quantity $Q$ for
particle motion. Therefore, the dynamics of the Hamiltonian $K$ in
Eq. (56) corresponding to the Lagrangian $\mathcal{L}_2$ of Eq.
(55) are integrable. No chaotic particle orbits but regular
Kolmogorov-Arnold-Moser (KAM) torus of particle orbits on the
Poincar\'{e} section $\theta=\pi/2$ with $p_\theta>0$ can be seen
in Fig. 1. This supports the dynamical integrability of
$\mathcal{L}_2$ to some extent. Two points are illustrated here.
Point 1: The integrability of a system cannot be established
numerically using Poincar\'{e} sections. In fact, numerical
methods can only demonstrate that regular dynamics occur for the
specific parameters and initial conditions under consideration,
but provide no information about whether regular dynamics exist
for other parameter sets and initial conditions. Integrability can
only be proven by finding a sufficient number of integrals of
motion. The integrable dynamics of Hamiltonian $K$ in Eq. (56) are
verified analytically via the existence of four conserved
quantities, namely $\mathcal{E}$, $\Phi$, $K$, and $Q$. Point 2:
The parameters used in Fig. 1 are chosen as $\omega =2$, $r_c
=29$, $K=-1/2$, $\mathcal{E}=1$, and $\Phi=1$. In addition, the
adopted numerical integration scheme for solving the Hamiltonian
$K$ of Eq. (56) is an eighth- and ninth-order Runge-Kutta-Fehlberg
algorithm [RKF8(9)] with variable step sizes. Although this
integrator shows secular drift in Hamiltonian errors, it can still
give reliable numerical results because of its high enough
accuracy.

The constant $Q$ of Eq. (57) is similar to the Carter constant,
which exists in the Kerr-Newman black hole spacetime with a
self-consistent electromagnetic potential [18,31]. It is also
present for extra source including magnetic charges,  cloud
strings and quintessence [27,51]. That is to say, the dynamics of
charged particles in these situations are integrable and regular.

\subsection{Chaotic dynamics of the Lagrangian $\mathcal{L}_1$ }

Based on Eq. (32) with Eqs. (52)-(54), the Lagrangian is rewritten
as
\begin{eqnarray}
\mathcal{L}_1 &=&
-\sqrt{1-\frac{2}{r}-\frac{\dot{r}^2}{1-\frac{2}{r}}-r^2\dot{\theta}^2
-r^2\sin^2\theta\dot{\phi}^2} \nonumber \\
&& -\frac{\omega}{r^2}[(r-r_c)^2 +4\theta^2].
\end{eqnarray}
Here, the Hamiltonian of Eq. (37) reads
\begin{eqnarray}
H_1 &=& E+\psi_1 \nonumber \\
&=& \sqrt{1-\frac{2}{r}}
\sqrt{1+(1-\frac{2}{r})p_r^2+\frac{p_\theta^2}{r^2}+\frac{L^2}{r^2\sin^2\theta}}
\nonumber \\
&& +\frac{\omega}{r^2}[(r-r_c)^2+4\theta^2],
\end{eqnarray}
where $E$ is an exact solution of the equation
\begin{eqnarray}
g^{tt}E^2+g^{rr}p^2_r+g^{\theta\theta}p^2_{\theta\theta}+g^{\phi\phi}L^2=-1.
\end{eqnarray}
This shows that the momenta $p_\mu$ obey the relation (36).
Hamiltonian $H_1$ has no separable form of the variables,
therefore, it is non-integrable. This result can be confirmed in
terms of numerical methods.

Let us select the vanishing angular momentum, $L=0$. All orbits of
massive particles are regular KAM torus for the energy
$\xi=H_1=9.979$ in Fig. 2(a). Seen from Eq. (59), a large value of
$\xi$ is allowed. As the energy increases to $\xi=10.03$, several
chaotic orbits occur in Fig. 2(b). There are more chaotic orbits
for $\xi=10.039$ in Fig. 2(c). These numerical results show that
the Hamiltonian $H_1$, i.e. the Lagrangian $\mathcal{L}_1$ of Eq.
(58), can allow for chaotic particle dynamics under some
circumstances. Chaos of a dynamical system must show
non-integrability of the dynamical system. Thus, the Lagrangian
$\mathcal{L}_1$ is non-integrable. On the other hand, nonzero
angular momentum with $L\neq0$ is also considered. For $L=6$,
chaos of massive particles does not emerge for $\xi=6$ in Fig.
3(a), but it does for $\xi=7.5$ and $\xi=8$ in Fig. 3 (b) and (c).
These demonstrations describe that chaos is allowed in the
Hamiltonian $H_1$ corresponding to the Lagrangian $\mathcal{L}_1$
for the case of nonzero angular momentum. This shows again the
non-integrability of $\mathcal{L}_1$.

As shown above, the dramatic difference in the dynamical
qualitative properties sufficiently proves the nonequivalence of
both Lagrangians $\mathcal{L}_1$ and $\mathcal{L}_2$ in Eqs. (55)
and (58).

\subsection{Extended external potentials}

If the external potential $\psi_1$ is extended to the form
\begin{eqnarray}
V=\psi_2=\frac{1}{r^2}[f_1(r)+f_2(\theta)],
\end{eqnarray}
where $f_1$ and $f_2$ are arbitrary continuously differentiable
functions, the fourth integral of motion similar to $Q$ in Eq.
(57) still exists. Therefore, the Hamiltonian $K$ of Eq. (56)
corresponding to the Lagrangian $\mathcal{L}_2$ exhibits
integrable and regular dynamics. However, it becomes
non-integrable and probably chaotic dynamics when the external
potential is given by $f_1$ plus $f_2$:
\begin{eqnarray}
V=\psi_3=f_1(r)+f_2(\theta),
\end{eqnarray}
or $f_1$ times $f_2$:
\begin{eqnarray}
V=\psi_4=f_1(r)f_2(\theta)~(f_1\neq\frac{1}{r^2}).
\end{eqnarray}

Regardless of whether the external potential is replaced with
$\psi_2$, $\psi_3$ or $\psi_4$, the Hamiltonian $H_1$ of Eq. (59)
corresponding to the Lagrangian $\mathcal{L}_1$ of Eq. (58) is
generally non-integrable and can allow chaos due to the difficulty
of separating variables leading to the absence of fourth integral
of motion. When the external potential is taken as
\begin{eqnarray}
V &=& \psi_5  = f_1(r) \nonumber \\
&&+f_2(r)\sqrt{1+(1-\frac{2}{r})p_r^2
+\frac{p_\theta^2}{r^2}+\frac{L^2}{r^2\sin^2\theta}},
\end{eqnarray}
the Hamilton-Jacobi equation of $H_1=E+\psi_5$ is separable and
therefore $\mathcal{L}_1$ shows integrable and regular dynamics.
However, the Hamilton-Jacobi equation of the Hamiltonian formalism
\begin{eqnarray}
K &=&
-\frac{\mathcal{E}^2}{2(1-\frac{2}{r})}+\frac{p^2_r}{2}(1-\frac{2}{r})
+ \frac{p^2_\theta}{2r^2} \nonumber \\
&& + \frac{\Phi^2}{2r^2\sin^2\theta}+\psi_5
\end{eqnarray}
is inseparable and therefore $\mathcal{L}_2$ shows non-integrable
and probably chaotic dynamics. For the external potential
$V=\psi_5$, the derivations of both Lagrangians $\mathcal{L}_1$
and $\mathcal{L}_2$ from the Hamiltonians $H_1$ and $K$ still rely
on the Legendre transformations:
$\mathcal{L}_1=\dot{x}^{i}p_i-H_1$,
$\mathcal{L}_2=\dot{x}^{\mu}P_\mu-K$, and the momenta
$\rightarrow$ the 4-velocities. If $f_2=0$ in the external
potential $ V=\psi_5$, then $\mathcal{L}_1$ and $\mathcal{L}_2$
are integrable and exhibit regular dynamics.

In a word, the above numerical tests and analytical discussions
sufficiently show that the two Lagrangian forms $\mathcal{L}_1$
and $\mathcal{L}_2$ have different dynamical behaviors for various
choices of the external potentials. In fact, $\mathcal{L}_1$ is
basically non-integrable for the external potentials excluding
$V=\psi_5$. Thus, chaos becomes possible. $\mathcal{L}_2$
possesses integrable dynamics for the cases of $V=\psi_2$
(including $V=\psi_1$) and $V=\psi_5$ with $f_2=0$, while
non-integrable and probably chaotic dynamics for the cases of
$V=\psi_3$, $V=\psi_4$ and $V=\psi_5$ with $f_2\neq0$. Obviously,
both Lagrangians are not dynamically equivalent.

\section{Remarks}\label{sec4}

As discussed extensively above, we have addressed the main
question of whether the two Lagrangians in Eqs. (3) and (4) are
dynamically equivalent. Now, we give several remarks to answer the
other questions mentioned in the Introduction Section.

\subsection{Default selection of $\mathcal{L}_1$ for generic external potentials}

It is highly probable that both Lagrangians  possess distinct
dynamic qualitative properties  for generic external potentials
unlike the electromagnetic potentials, as shown by the above
theoretical and numerical results. Which of them is plausible? Of
course, $\mathcal{L}_1$ is. The fundamental reason is that the
Hamiltonian of $\mathcal{L}_1$ inherently incorporates the mass
shell constraint, while that of $\mathcal{L}_2$ does not. As a
result, nonphysical orbital dynamics including chaos can be
permitted in $\mathcal{L}_2$. For example, $\mathcal{L}_2$ has
nonphysical integrable and regular dynamics for the external
potentials $V=\psi_2$ (including $V=\psi_1$) and $V=\psi_5$ with
$f_2=0$, but spurious non-integrable and chaotic dynamics for the
external potentials $V=\psi_3$, $V=\psi_4$ and $V=\psi_5$ with
$f_2\neq0$.

As stated in the Introduction Section, $\mathcal{L}_1$ possesses
theoretical superiority over $\mathcal{L}_2$. It serves as the
universal form for describing massive particle motion in weak or
strong gravitational field problems involving gravitational and
matter fields. It is applicable to non-relativistic, relativistic,
and ultrarelativistic regimes. Thus, it is highly recommended in
general.

\subsection{Validity of $\mathcal{L}_1$ for massless particle motion}

Which of the two Lagrangians is valid to describe the motion of
massless particles (including photons)? We discuss this question
in two cases.

$\mathcal{L}_1$ and $\mathcal{L}_2$ of Eqs. (1) and (2) seem to be
useless for the massless particle motion with $m=0$. This is
because $\mathcal{L}_1$ and $\mathcal{L}_2$ are identically equal
to zero without doubt if $m=0$ is directly substituted into Eqs.
(1) and (2). When the condition $m=0$ is not considered, we derive
the  Hamiltonians $H_1$ and $H_2$ of Eqs. (37) and (43) from both
Lagrangians. Now, the circumstances of $m=\psi=0$ are given in
$H_1$ and $H_2$ of Eqs. (37) and (43). As a result, $H_1$ and
$H_2$ have the same expressions and are suitable for the
description of massless particle dynamics. Although
$\mathcal{L}_2$ does not embody the mass shell constraint
immanently, it does when the additional constraint (23) is given
to the Hamiltonian of $\mathcal{L}_2$. This result is based on no
matter corresponding to the vanishing external potential.

When the massless particle interacts with matter described by the
external potential $V=\psi\neq0$, the Hamiltonian $H_2$ of Eq.
(43) does not contain the external potential $\psi$. Therefore,
$\mathcal{L}_2$ becomes useless for massless particle motion.
However, the Hamiltonian $H_1$ of Eq. (37) with the external
potential $\psi$, i.e. $\mathcal{L}_1$, can validly work. Because
of this point, $\mathcal{L}_1$ is used to explore the motion of
massless particles in Refs. [15-17,56,57]. Although the Lagrangian
formalism $\mathcal{L}_1$ is not explicitly used in Refs.
[15-17,56,57], the Hamiltonian like $H_1$ of Eq. (37) comes from
the third path for obtaining the energy $E$ of the mass shell
constraint (36) for the Lagrange $\mathcal{L}_1$. Eq. (5) in Ref.
[15],  Eq. (8) in Ref. [16], Eq. (7) in Ref. [17], Eq. (14) of
Ref. [56], and Eq. (2.6) of Ref. [57] are the energies $E$ solved
from the mass shell constraints. Thus, Refs. [15-17,56,57] are
attributed to the application of the Lagrangian $\mathcal{L}_1$ to
the motion of massless particles.

Square-root Lagrangian $\mathcal{L}_1$ is suitably applicable to
massless particle  motion regardless of whether the external
potential is vanishing or not.

\subsection{Suitability of  $\mathcal{L}_2$ for electromagnetic potentials}

Non-square-root Lagrangian $\mathcal{L}_2$ is generally applied to
low-energy weak fields and non-relativistic limits. It does not
self-consistently contain the mass shell constraint in general
even if the additional constraint (23) is given to its
Hamiltonian.

Is its application to strong gravitational field problems
universally nonphysical? This is not always true and there are
exceptions. It is theoretically perfect from relativistic physics
that many authors [18-39,58,59] applied $\mathcal{L}_2$ (i.e. the
Hamiltonian formalism $K$ of Eq. (22) with the additional
constraint of Eq. (23)) to investigate the dynamics of charged
particles near black holes with external electromagnetic fields
represented by electromagnetic potentials. This is because both
$\mathcal{L}_1$ and $\mathcal{L}_2$ have dynamical equivalence
when the Hamiltonian $K$ (22) of $\mathcal{L}_2$ is specified by
the additional condition (23). In this case, $\mathcal{L}_2$ like
$\mathcal{L}_1$ also accurately satisfies the mass shell
constraint condition (24) and strictly obeys the theory of general
relativity. $\mathcal{L}_2$ has some advantages in numerical
simulations because of simplifying numerical integrations and
reducing computational cost. In particular, the Hamiltonian
formalism $K$ (22) with the additional constraint (23) is easy to
programmatically implement explicit symplectic integrators with
high computational efficiency [48-50,52-54,59]. These symplectic
methods are very appropriate for tracking chaotic behavior of
charged particle orbits around black holes in external
electromagnetic fields.

Due to the convenience of computations, $\mathcal{L}_2$ of Eq.
(15) rather than $\mathcal{L}_1$ of Eq. (14) is preferred for the
study of charged particle dynamics in the combined gravitational
and electromagnetic fields. Above all, $\mathcal{L}_2$
corresponding to the Hamiltonian formalism $K$ (22) with the
additional constraint (23) contains the consistency of the mass
shell constraint, as $\mathcal{L}_1$ does.

These remarks address the practical choice between $\mathcal{L}_1$
and $\mathcal{L}_2$. They further confirm the nonequivalence of
the two Lagrangians in Eqs. (3) and (4) for non-electromagnetic
external potentials. The equivalence of these Lagrangians was
established only in Footnote 1 and was not discussed elsewhere in
Ref. [13]. Lyapunov exponents were calculated for both Lagrangians
under an electromagnetic external potential, while chaotic
dynamics was examined in $\mathcal{L}_1$ (but not in
$\mathcal{L}_2$) for an external potential analogous to the
harmonic oscillator potential. Accordingly, both the methodology
and results presented in Ref. [13] are undoubtedly correct.

\section{Summary} \label{sec5}

The two Lagrangian formalisms $\mathcal{L}_1$ and $\mathcal{L}_2$
in Eqs. (3) and (4) are widely employed to describe particle
motion in gravitational and matter fields. We consider a static,
axisymmetric four-dimensional metric $g_{\mu\nu}$ with $g_{0i}=0$.
The metric is coupled to an axisymmetric external matter potential
$V$ that depends on the spacetime coordinates. Under these
assumptions, the particle energy and angular momentum are
conserved in both Lagrangian systems. The properties and relations
of these two Lagrangians are then analyzed.

In the absence of an external potential, the square-root
Lagrangian $\mathcal{L}_1$ follows from the principle of least
action. By contrast, the alternative Lagrangian $\mathcal{L}_2$
does not contain a square-root structure. The square-root
Lagrangian $\mathcal{L}_1$ naturally incorporates the mass shell
constraint, which is directly consistent with a constrained
Hamiltonian. The non-square-root Lagrangian $\mathcal{L}_2$  can
also satisfy the mass shell constraint when an additional
constraint is imposed on its Hamiltonian. In fact, the two
Hamiltonian formalisms are equivalent. Hence, the two Lagrangian
formulations are undoubtedly dynamically equivalent.

For the case of an electromagnetic external potential, the
redefined momentum associated with the square-root Lagrangian
$\mathcal{L}_1$ inherently and exactly fulfills the redefined mass
shell constraint, which is in agreement with a constrained
Hamiltonian description. This mass shell constraint also emerges
for the non-square-root Lagrangian $\mathcal{L}_2$ when a suitable
additional constraint is imposed on its Hamiltonian. The two
Hamiltonian formalisms coincide exactly. This demonstrates the
dynamical equivalence of both Lagrangian formulations.

For a non-electromagnetic external potential such as $V=\psi_2$,
the Hamiltonian formulations of $\mathcal{L}_1$ and
$\mathcal{L}_2$ differ. The Hamiltonian of $\mathcal{L}_1$
intrinsically satisfies the mass shell constraint, whereas that of
$\mathcal{L}_2$ does not. This explicitly demonstrates that
$\mathcal{L}_1$ and $\mathcal{L}_2$ correspond to two distinct
dynamical systems. To illustrate their different dynamical
behaviors, we adopt the Schwarzschild metric and an artificially
constructed mechanical potential as a toy model. Numerical results
show that the first Lagrangian exhibits non-integrable and chaotic
dynamics owing to the absence of a fourth constant of motion. In
contrast, our analysis reveals that the second Lagrangian always
yields integrable and regular dynamics due to the presence of a
fourth constant of motion. When the external potential is
generalized to the more general form $V=\psi_5$ with $f_2\neq0$,
$\mathcal{L}_1$ remains integrable and regular, while
$\mathcal{L}_2$ becomes non-integrable and may admit chaotic
behavior. Accordingly, the two Lagrangians are not dynamically
equivalent.

In summary, the dynamical equivalence of the two Lagrangian
formalisms depends crucially on the choice of external potential.
They are dynamically equivalent when the external potential
vanishes or takes the form of an electromagnetic potential. In
these cases, the particle energies derived from the mass shell
constraints can be directly adopted as the Hamiltonians of the
Lagrangian systems, which consistently satisfy the mass shell
conditions. However, the two Lagrangians are generally not
dynamically equivalent for other types of external potentials. For
the first Lagrangian, the Hamiltonian is constructed from the
energy that fulfills the mass shell constraint, whereas for the
second Lagrangian, the corresponding energy does not satisfy the
mass shell constraint. In short, the first Lagrangian preserves
the mass shell constraint, while the second one does not.

In practical applications, the first Lagrangian is generally
highly recommended, as it is theoretically superior to the second
one in most cases. It provides a universal description for the
motion of both massive and massless particles in both strong and
weak gravitational fields. By contrast, the second Lagrangian is
typically suitable for low-energy, weak-field regimes and
non-relativistic limits. However, for the dynamics of charged
particles near black holes in the presence of external
electromagnetic fields, the second Lagrangian corresponding to the
Hamiltonian $K$ of Eq. (22) with the additional constraint of Eq.
(23) is preferred over the first one, which corresponds to the
Hamiltonian $H_1$ of Eq. (19). This is because the Hamiltonian $K$
of Eq. (22) supplemented by constraint of Eq. (23) is equivalent
to $H_1$ of Eq. (19) and accurately satisfies the mass shell
constraint. Furthermore, both the mathematical tractability and
computational efficiency of the Hamiltonian associated with the
second Lagrangian are superior to those of the first Lagrangian.

\textbf{Acknowledgements}: The authors would like to thank
Professor Xian-Hui Ge for helpful discussions. They are very
grateful to a referee for valuable comments and suggestions, too.
This research is supported by the National Natural Science
Foundation of China (Grant No. 12573077).

\textbf{Data Availability Statement} This manuscript has no
associated data or the data will not be deposited. [Author's
comment: The datasets generated and/or analysed during the current
study are available from the corresponding author on reasonable
request.]

\textbf{Code Availability Statement} Code/software will be made
available on reasonable request. [Author's comment: The
code/software generated during and/or analysed during the current
study is available from the corresponding author on reasonable
request.]

\textbf{Conflicts of Interest}: The authors declare no conflict of
interest.

\begin{figure*}[htpb]
        \centering{
        \includegraphics[width=20pc]{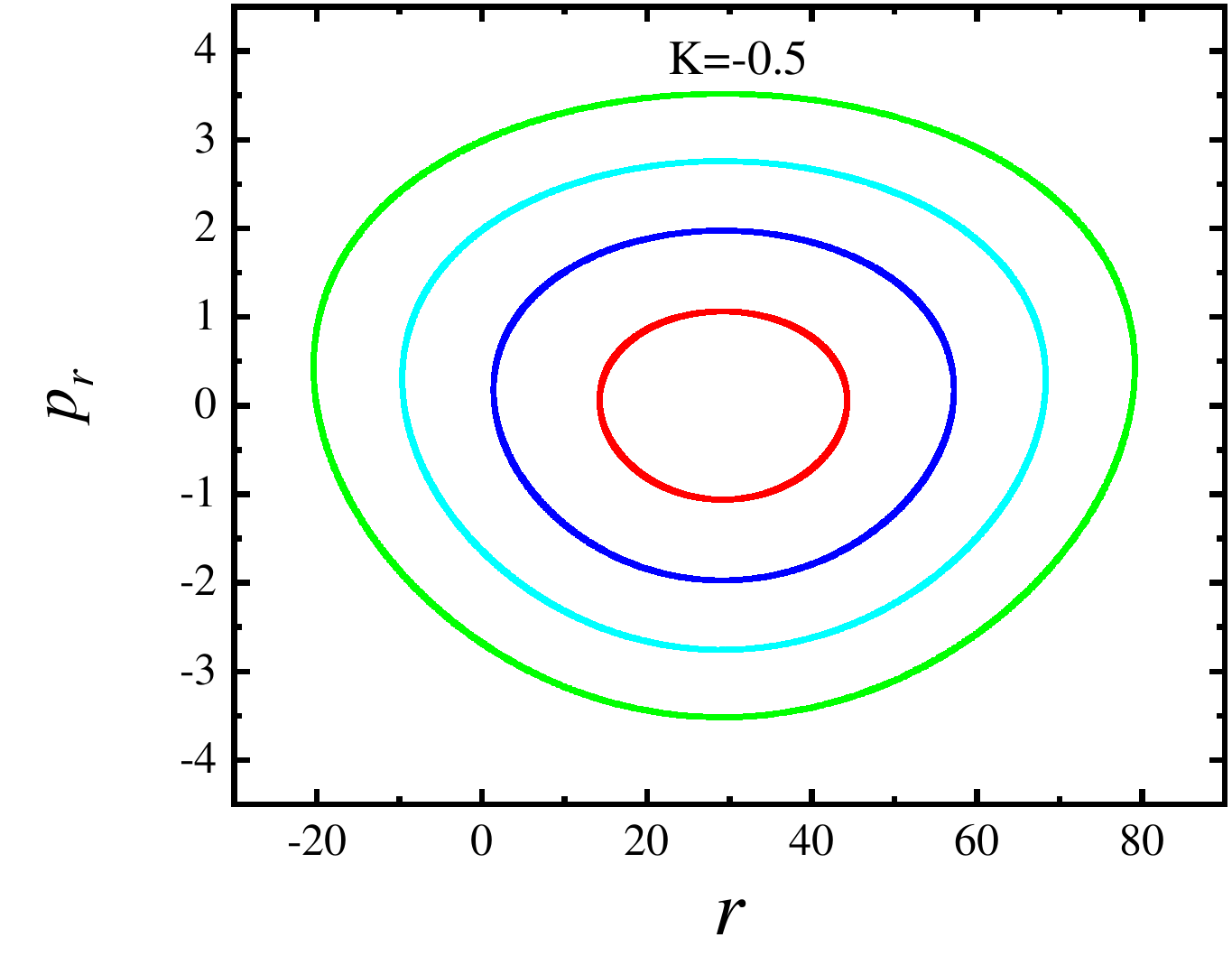}
\caption{Poincar\'{e} section at the plane $\theta=\pi/2$ with
$p_\theta>0$. The motion of particles is considered in the
Hamiltonian $K$ of Eq. (56) corresponding to the Lagrangian
$\mathcal{L}_2$ of Eq. (55). The parameters are $\omega =2$, $r_c
=29$, $K=-1/2$ and $\Phi=1$. All the plotted particle orbits are
regular KAM torus. }}
\end{figure*}

\begin{figure*}[htpb]
        \centering{
        \includegraphics[width=12pc]{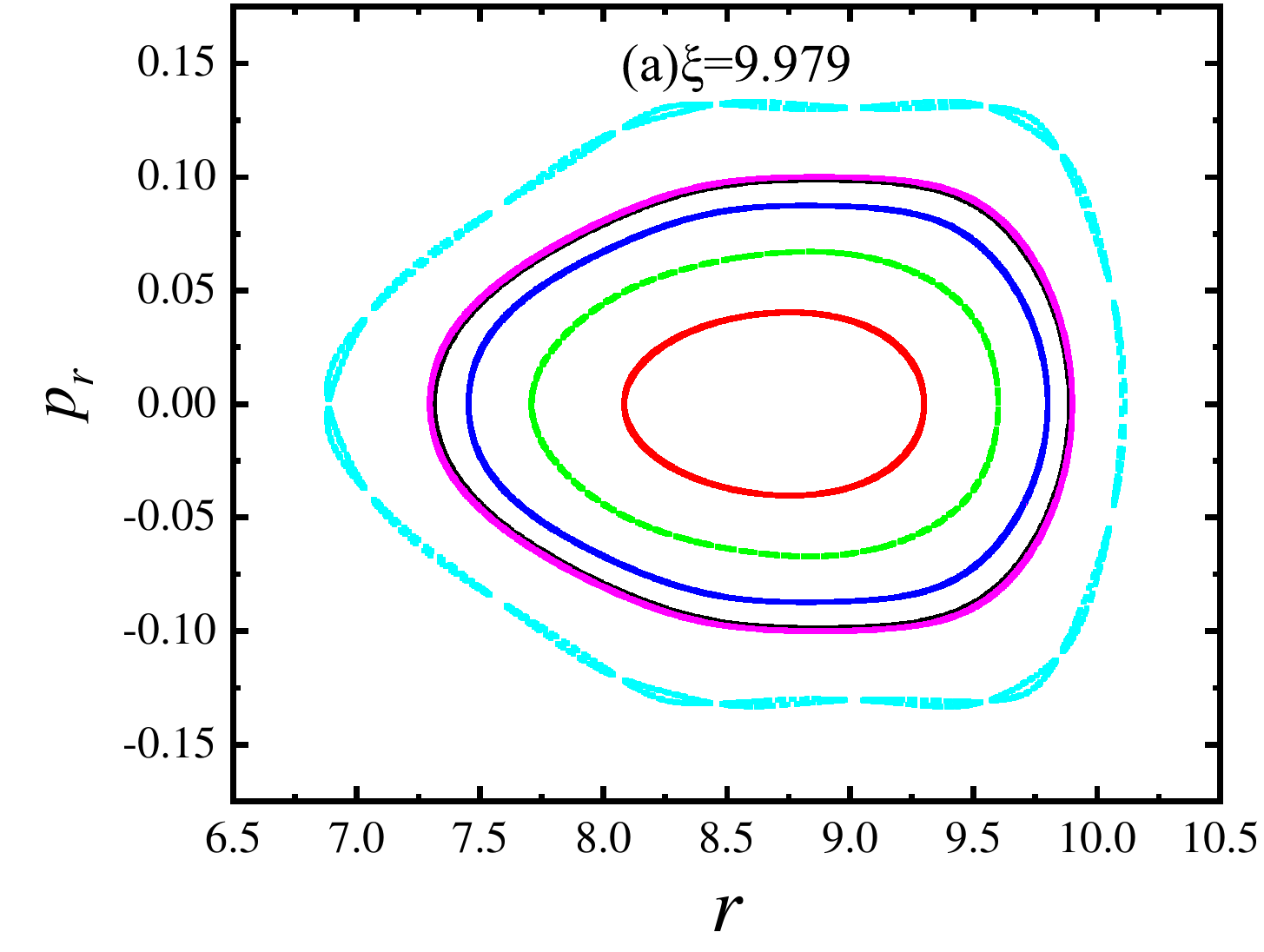}
        \includegraphics[width=12pc]{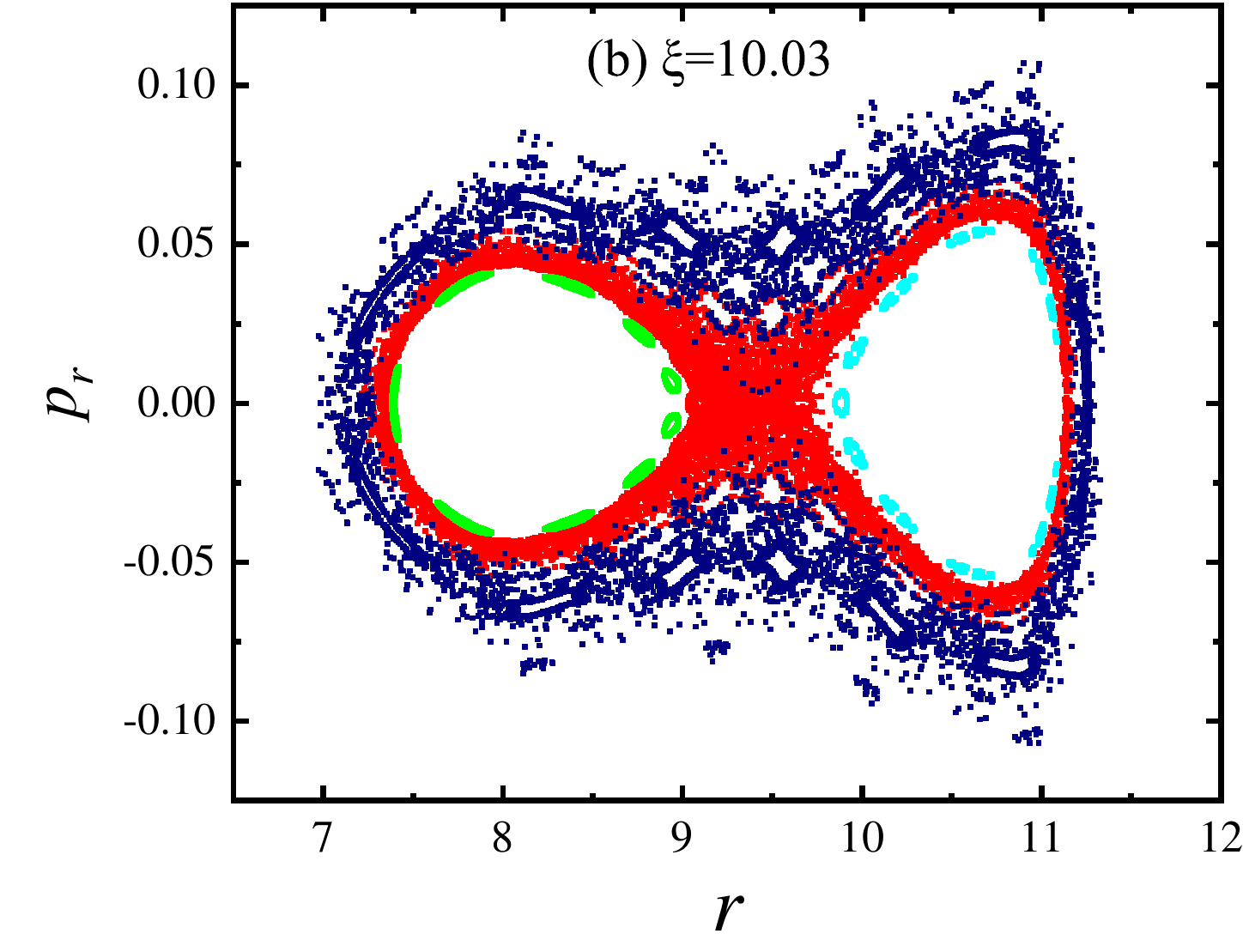}
        \includegraphics[width=12pc]{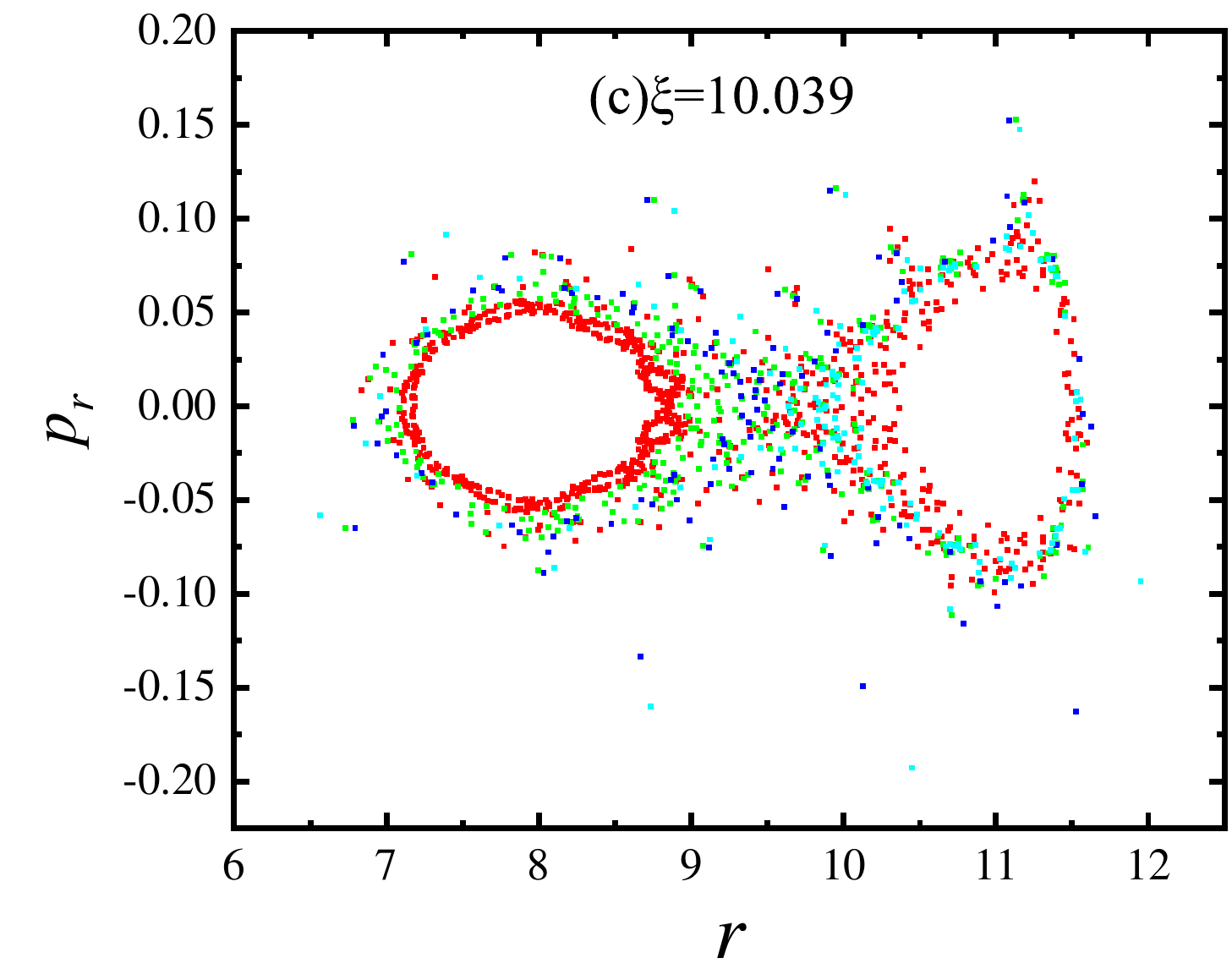}
\caption{Poincar\'{e} sections for three values of the energy
$\xi=H_1$ and the vanishing angular momentum $L=0$. The motion of
particles is considered in the Hamiltonian $H_1$ of Eq. (59),
which corresponds to the Lagrangian $\mathcal{L}_1$ of Eq. (58).
The parameters are $\omega =10$ and $r_c =1$. Chaos does not exist
for (a) $\xi=9.979$, but does for (b) $\xi=10.03$ and (c)
$\xi=10.039$. The onset of chaos sufficiently shows the
non-integrability of the Lagrangian $\mathcal{L}_1$. } }
\end{figure*}

\begin{figure*}[htpb]
    \centering{
    \includegraphics[width=12pc]{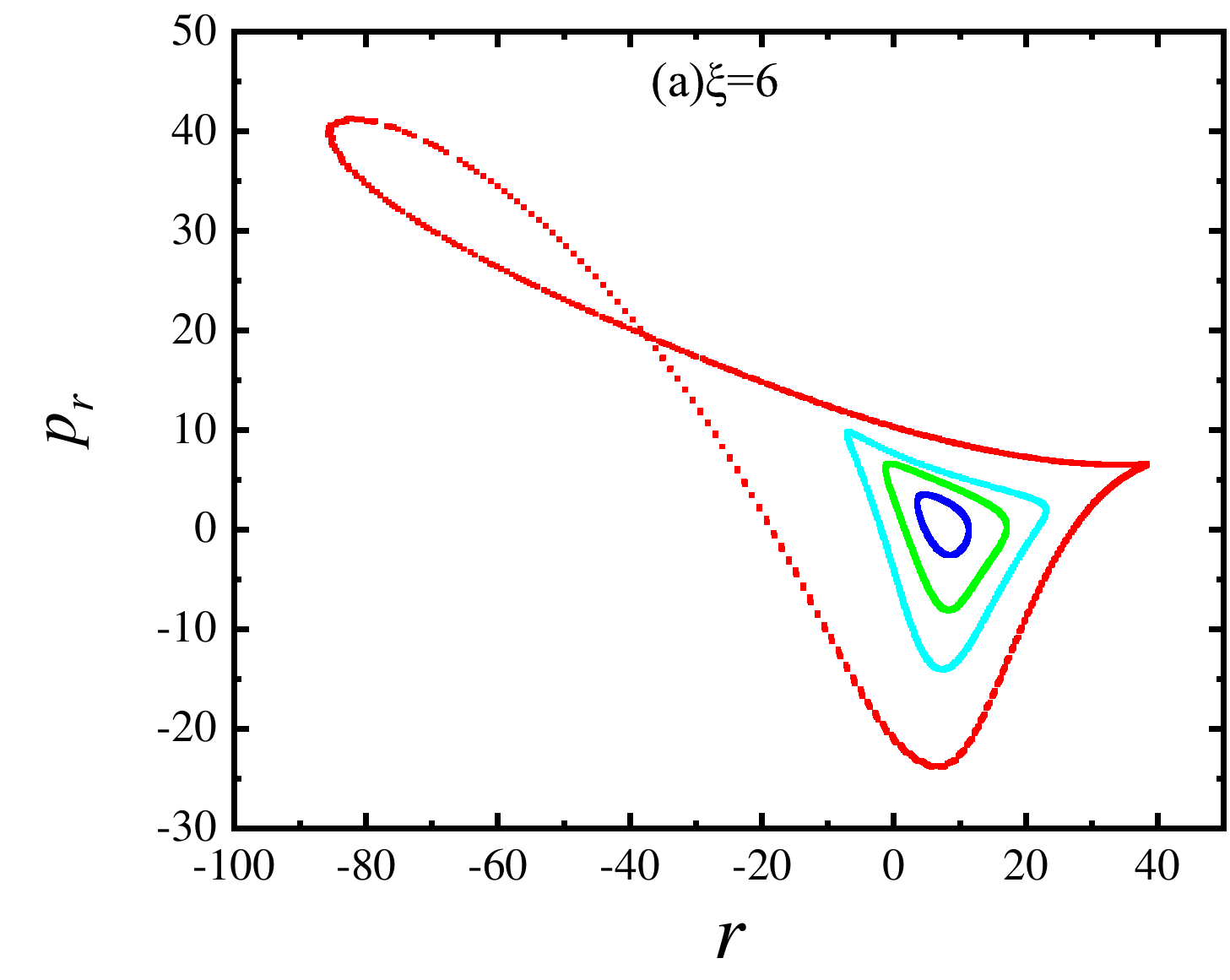}
    \includegraphics[width=12pc]{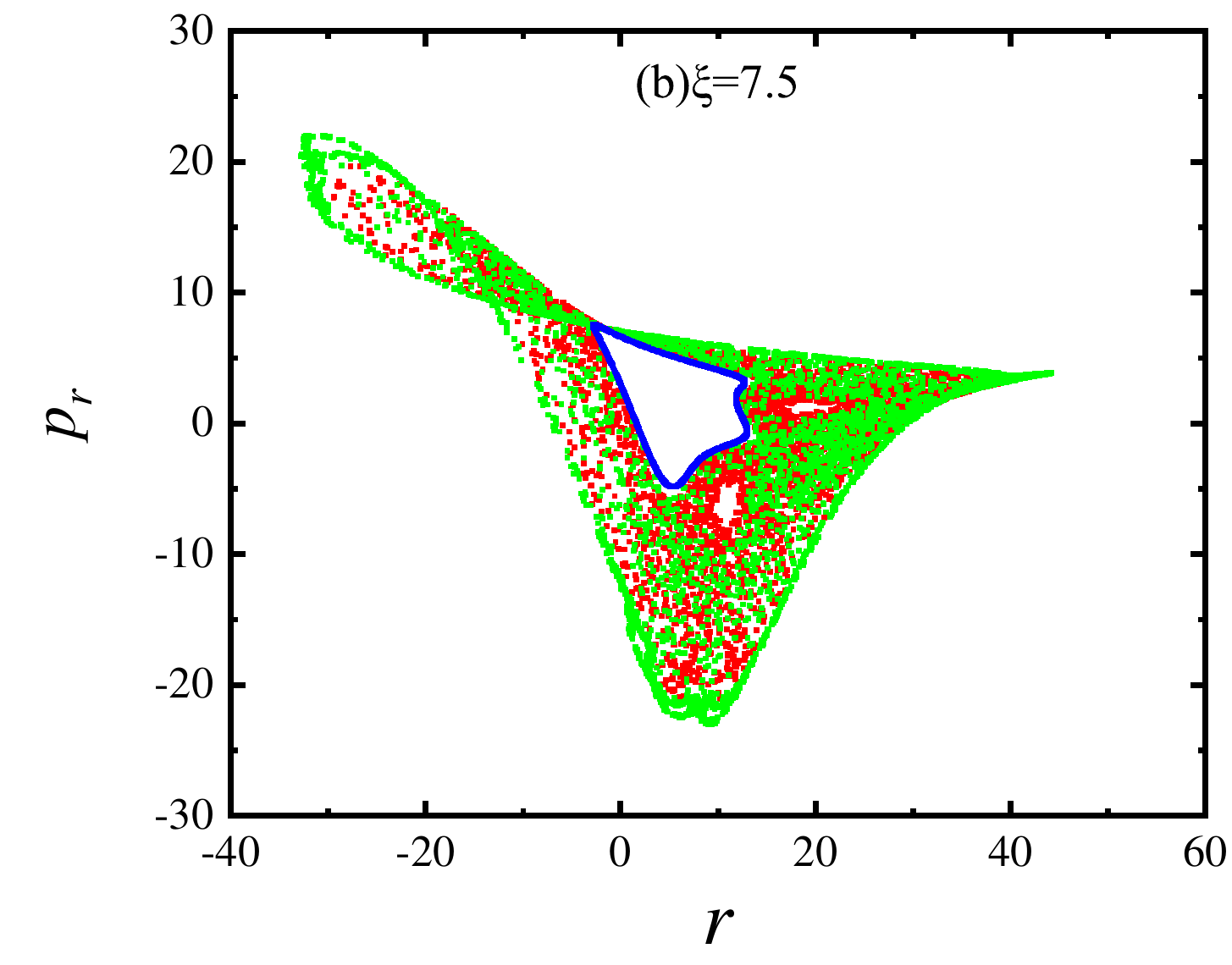}
    \includegraphics[width=12pc]{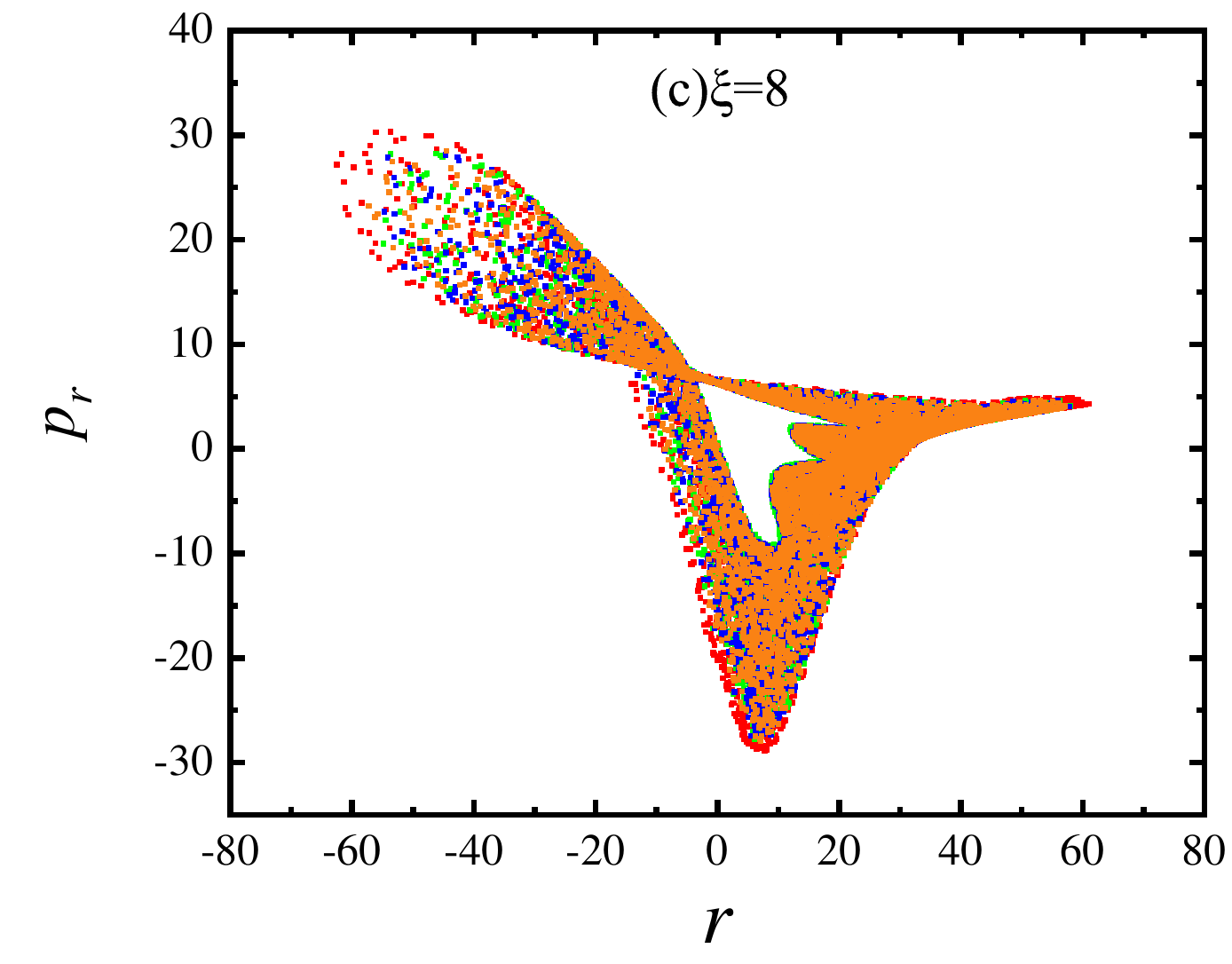}
\caption{Same as Fig. 2 but for the non-vanishing angular momentum
$L=6$. The parameters are $\omega =10$ and $r_c =7$. Chaos does
not occur for (a) $\xi=6$, but does for (b) $\xi=7.5$ and (c)
$\xi=8$. The onset of chaos powerfully supports the
non-integrability of the Lagrangian $\mathcal{L}_1$. }}
\end{figure*}

\end{document}